\documentclass[aps,reprint,groupedaddress,notitlepage,amsmath,amssymb,amsfonts,bm,subfigure]{revtex4-2}

\usepackage[utf8]{inputenc}
\usepackage[T1]{fontenc}

\usepackage[american]{babel}
\usepackage[autostyle=true]{csquotes}
\usepackage{graphicx,bm}

\usepackage{xcolor}
\usepackage[normalem]{ulem}

\usepackage[colorlinks=true]{hyperref}
\usepackage[sort&compress, capitalize]{cleveref}
\newcommand{\norm}[1]{\left\lVert#1\right\rVert}

\DeclareMathOperator*{\argmin}{arg\,min}

\usepackage[normalem]{ulem}
\usepackage{color}

\usepackage{graphicx,bm}
\usepackage{epstopdf, epsfig}

\begin{document}

%%%% Article title to be placed here

\title[Model of edge collapse]{A model for the collapse of the edge when two transitions routes compete}

\author{Miguel Beneitez$^1$}
\email{beneitez@kth.se}

\author{Yohann Duguet$^2$}
\email{duguet@limsi.fr}

\author{Dan S. Henningson$^1$}
\email{henning@kth.se}
\affiliation{$^1$Linné FLOW Centre and Swedish e-Science Research Centre (SeRC), KTH Engineering Mechanics, Royal Institute of Technology, SE-10044 Stockholm, Sweden\\
$^2$LIMSI-CNRS, Univ. Paris-Saclay, P91405 Orsay, France}
\date{\today}

\begin{abstract}
The transition to turbulence in many shear flows proceeds along two competing routes, one linked with finite-amplitude disturbances and the other one originating from a linear instability, as in e.g. boundary layer flows. The dynamical systems concept of edge manifold has been suggested in the subcritical case to explain the partition of the state space of the system. This investigation is devoted to the evolution of the edge manifold when a linear stability is added in such subcritical systems, a situation poorly studied despite its prevalence in realistic fluid flows. In particular the fate of the edge state as a mediator of transition is unclear. A deterministic three-dimensional model is suggested, parametrised by the linear instability growth rate. The edge manifold  evolves topologically, via a global saddle-loop bifurcation, from the separatrix between two attraction basins to the mediator between two transition routes. For larger instability rates, the stable manifold of the saddle point increases in codimension from 1 to 2 after an additional local saddle node bifurcation, causing the collapse of the edge manifold. As the growth rate is increased, three different regimes of this model are identified, each one associated with a flow case from the recent hydrodynamic literature. A simple nonautonomous generalisation of the model is also suggested in order to capture the complexity of spatially developing flows.
\end{abstract}

\maketitle

\section{Introduction}

After more than a century of both theoretical, experimental and computational progress, predicting the transition from laminar to turbulence in usual fluid flows is still a puzzle for mathematicians, physicists and engineers \cite{reynolds1883,eckhardt2007turbulence}. The situation appears even more complex when different routes towards the same turbulent state are in competition. The theory of dynamical systems is an elegant way to rationalize such situations because it provides a cartography of the space of possible initial conditions depending on their outcome at a later time. The key concepts here are those of basins of attraction and boundaries separating disjoint basins of attraction \cite{nusse1996basins}. For purely supercritical systems such as co-rotating Taylor-Couette flow or the Rayleigh-B\'enard set-up, transition is triggered by any infinitesimal disturbance provided the control parameter (usually the Reynolds number or the Rayleigh number) exceeds a given threshold \cite{drazin2004hydrodynamic}. In such a case, the whole state space becomes the attraction basin of the turbulent state. For classical subcritical systems such as pipe or plane Couette flow, the state space is usually partitioned into two basins of attraction, one for the laminar state and the other one for the turbulent one \cite{itano2001dynamics,eckhardt2007turbulence}. When both the laminar and the turbulent regime are attracting sets (in the sense that trajectories stay in their neighbourhood for arbitrary large times), the basin boundary common to these two basins is a smooth hypersurface called \emph{the edge} \cite{Skufca2006}. The edge is an invariant set for the flow dynamics, and it is of codimension 1. In the simplest case, it is the stable manifold of a given unstable state called \emph{the edge state}, a relative attractor within that manifold \cite{duguet2008transition}. The most popular method to identify the edge state(s) of the system relies on bisection \cite{itano2001dynamics,schneider2008edge}. Alternatives to this approach have been suggested but usually rely on some pre-knowledge of the dynamical nature of the edge state \citep{willis2017surfing,linkmann2019dynamic,beneitez2020lcs}. The study of the edge manifold becomes mathematically more complex in a situation relevant for the lowest Reynolds numbers close to the onset of transition: the turbulent set, although still a chaotic set, is no longer an attractor. Trajectories visiting it have a finite probability to reach the laminar state after an arbitrary long transient time \cite{sweet2001stagger,Skufca2006,hof2006finite}. This has lead to a generalisation of the concept of edge manifold, seen as a boundary between two types of a trajectories that happen to have the same asymptotic outcome, namely the laminar regime. Ref.\  \cite{lebovitz2012boundary} introduced a distinction, depending on whether the edge manifold splits the state space into two disjoint attractions basins (hard type) or not (soft type).
\begin{figure*}
    \centering
    \includegraphics[width=0.46\textwidth]{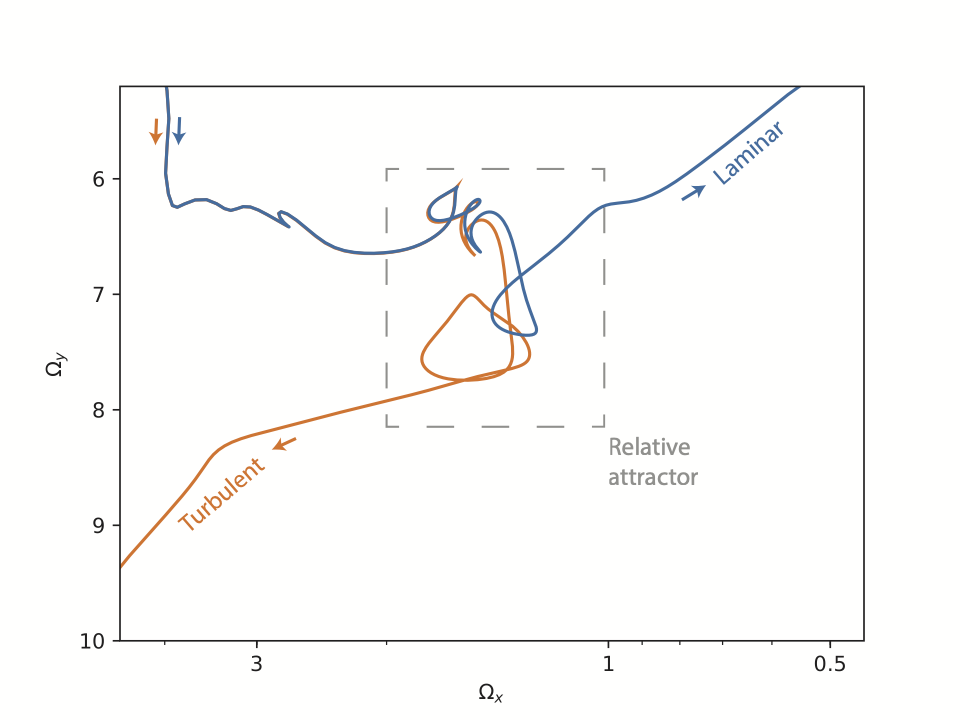}\put(-230,165){$(a)$}\put(-215,160){$\times 10^{-4}$}
    \includegraphics[width=0.46\textwidth]{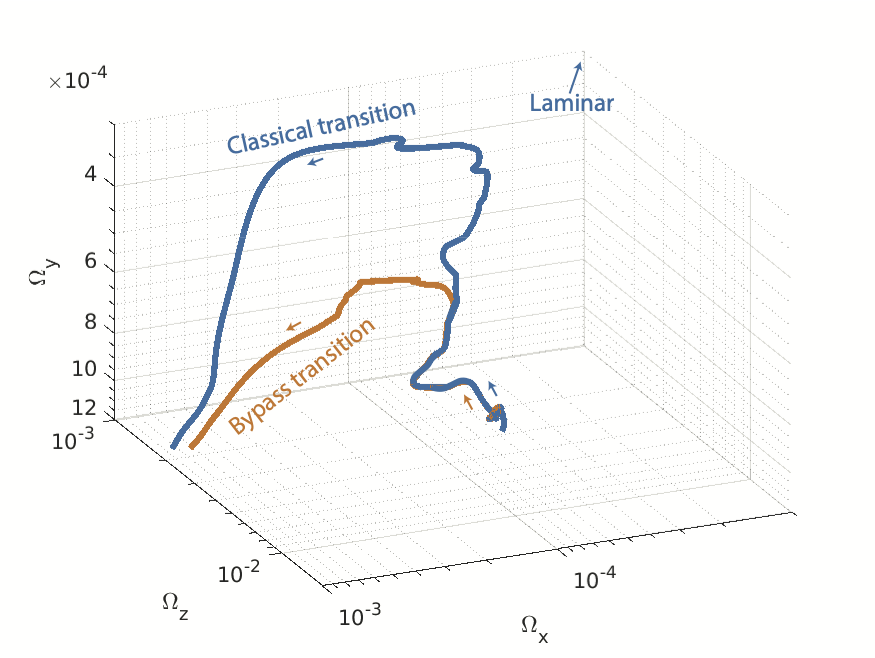}\put(-230,165){$(b)$}\put(-255,21){$\times 10^{-4}$}\\
    \includegraphics[width=0.85\textwidth]{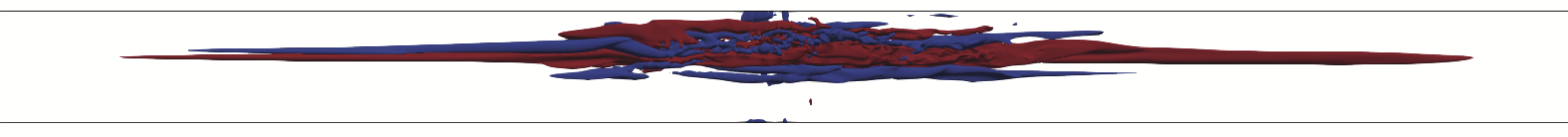}\put(-460,40){$(c)$}\\
    \includegraphics[width=0.60\textwidth]{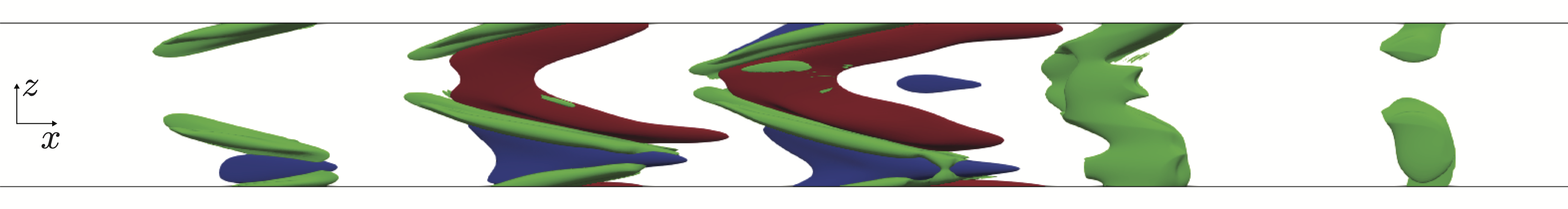}\put(-395,45){$(d)$}
    \caption{Edge tracking in Blasius boundary layer (see \cite{beneitez2019edge} for a detailed study). (a) State portrait at moderate times $t \approx [0,5000]$ using the vorticity variables $\Omega_x$ and $\Omega_y$ defined in \cite{Duguet2012,beneitez2019edge}. (b) State portrait at large times $t \approx [700,14000]$ using the variables $\Omega_x$, $\Omega_y$ and $\Omega_z$. Trajectories correspond to the classical transition route approaching the laminar state (blue) and the bypass transition route (orange). (c) and (d) $xz$-view of three-dimensional perspective along two different coexisiting transition routes. Contours of $\lambda_2=-9\times 10^{-6}$ (green), isosurfaces of streamwise perturbation velocity with respect to the spanwise mean with values $=0.06$ and $-0.08$ (red and blue respectively), flow from left to right. (c) Bypass transition (d) Classical transition.}
    \label{fig:blasius}
\end{figure*}

Conceptually, there is a symmetric configuration which has been little studied so far despite its relevance to transitional flows at high Reynolds numbers: what happens to the edge manifold in the case where the laminar state loses its laminar stability (while the turbulent state remains attracting)? Such a linear instability of the laminar state in an otherwise subcritical fluid system occurs at least in three of the most important examples of shear flow transition: the flow inside a curved pipe \cite{kuhnen2015subcritical}, inside a parallel channel \cite{orszag1971accurate}, and the Blasius boundary layer flow developing over a flat plate \cite{schmid2001stability}. In these cases the subcritical nature of the transition (the bypass transition scenario) is in competition with the exponential growth, followed by their destabilisation, of the so-called Tollmien-Schlichting (TS) waves (the classical transition scenario) \cite{morkovin1969many}. Even in the absence of such well understood linear instability, other general destabilisation mechanisms can be in competition with the usual bypass picture: presence of small roughness at the walls \cite{bucci2018roughness}, geometrical defect to the base flow \cite{bottaro2003effect} or, among other possibilities, competing instability mechanisms due to additional parameters \cite{clever_busse_1992}. Several important fundamental questions arise in such cases:
what happens to the basin boundary and to its role as a mediator of transition \cite{khapko2016edge}? Can edge states still be identified and what is their role in the state space? Does the geometric notion of edge manifold at least make sense mathematically speaking? Is there any way to select or control a given transition route at the expense of the other one? Can low-order models faithfully reproduce the complex dynamics at play, and if so, to which extent?

The three canonical cases of curved pipe flow, plane channel flow and the Blasius boundary layer have been assessed numerically in very recent investigations. For curved pipe flow finite curvature leads, above some threshold in the Reynolds number, to an additional instability absent from the straight pipe case \cite{canton2020critical}. This instability leads to a limit cycle replacing the laminar state. The edge manifold  generalises hence to the separatrix between the turbulent state and a new attractor replacing dynamically the traditional laminar fixed point. In plane Poiseuille flow, for the parameters chosen in the corresponding study, the edge state is a travelling wave solution appearing in a saddle-node bifurcation at a finite Reynolds number $R\approx 459$, while the instability of the base flow to TS waves does not occur before $R=5815$ \cite{zammert2019transition}. As $R$ is increased from low to high values, the classical route emerges in parallel to the bypass route, and concerns an increasingly large subset of the state space of initial conditions \cite{zammert2019transition}. However for $R = 5855$ it is reported that the edge state still exists as a finite-amplitude travelling wave, whose stable manifold still separates the initial conditions leading to the turbulent state by involving TS waves from those who lead to transition without TS waves taking an active role. The case of the Blasius boundary layer, despite its relevance to transition in wind tunnels, presents additional difficulties for bifurcation studies: because of the spatial development of the laminar base flow in the downstream direction, there is no independent control parameter. A Reynolds number can be constructed as in most flows, however it should rather be interpreted as a spatial coordinate \cite{schmid2001stability}. There, the bisection algorithm has also been employed in order to determine an edge trajectory converging neither to the laminar nor to  the turbulent state. The computation cost of the whole approach, linked to the cost of simulating a spatially extended three-dimensional flow field, made conclusions ambiguous until a recent time. The bisection method successfully identifies a recognisable spatially localised coherent structure over moderate finite times in the form of a long velocity streak \cite{Cherubini_2011,Duguet2012}. However, for even larger time horizon time it becomes increasingly ambiguous to label a given trajectory as transitioning via the bypass or classical route \cite{beneitez2019edge}. This results in a failure of the standard bisection algorithm for the asymptotic edge regime. An illustration of the two different routes to turbulence starting from nearby initial conditions is shown in Fig.\ \ref{fig:blasius} using a physical space visualisation\footnote{In physical space the coordinates $x,y,z$ correspond to the streamwise, wall-normal and spanwise directions. $\lambda_2$ is the criterion for vortex identification introduced in \cite{jeong_hussain_1995}.} and state portraits. The state space visualisations in Fig.\ \ref{fig:blasius}(a),(b) show that despite a common initial path, the two trajectories separate rapidly but converge later towards the same turbulent regime. The physical space visualisations in Fig.\ \ref{fig:blasius}(c),(d) highlight the very different coherent structures present along either route: elongated in the streamwise direction for the bypass route, in the spanwise direction for the classical route.
The three present examples, based on realistic Navier-Stokes simulations, correspond to three scenarios distinct from the classical bistable picture presented in \cite{itano2001dynamics}. This wealth of behaviours suggests that a better understanding of the fate of the edge manifold is welcome as soon as a linear instability competes with an already existing subcritical transition picture. Low-dimensional models of subcritical transition have proven a strong companion tool for an improved understanding of dynamical features such as the role of the edge in transition \cite{Dauchot_1997,waleffe1997self,cossu2005non,vollmer2009basin,lebovitz2013edge,pausch2015direct}. In this tradition, we suggest here a novel nonlinear model of subcritical transition, inspired by the two-dimensional model in Ref.\ \cite{Dauchot_1997} to which a linear instability mechanism is added consistently with all the hydrodynamic constraints. As we shall see, the suggested autonomous model parametrised by the unstable growth rate displays three parameter regions of interest. In region $I$, several local bifurcations change the attracting laminar state without altering the global structure of the state space, a dynamical regime qualitatively akin to  Ref.\ \cite{canton2020critical}. In region $II$, the state space structure is globally modified after a global bifurcation, only one basin of attraction remains while the edge state still exists, similarly to Ref.\ \cite{zammert2019transition}. In region $III$ the edge state ceases to exist and the edge collapses, qualitatively closer to the dynamics reported in Ref.\ \cite{beneitez2019edge}. The study of these regions, together with an extension to a nonautonomous system allows for an explanation of all the dynamical regimes displayed by the previous fluid flow examples and suggests new directions.

The issues raised in this article go easily beyond the realm of hydrodynamic transition to turbulence. Any nonlinear dissipative system with two competing attracting regimes is concerned with the notion of basin boundary and edge manifold. At the level of modelling, many of the low-order models from mechanics show strong similarities with the models used in hydrodynamics \cite{strogatz2001nonlinear,marquardt2006dynamical,nave2019global,beneitez2020lcs}. The generalisation of the edge concept suggested using the present model is hence of interest for many dissipative bistable systems.

The paper is structured as follows: the reference two-dimensional Dauchot-Manneville system is introduced together with its autonomous three-dimensional generalisation in Section \ref{sec:DM2D} and \ref{sec:DM3D}, respectively. The parametric analysis of the 3D model is carried out in Section \ref{sec:parStudy}, \ref{sec:globalbif} and \ref{sec:accessibility}. A nonautonomous version of the model is introduced and analysed in Section \ref{sec:DM3Dnonauto}. Eventually, the relevance of the two models to the original hydrodynamical context is discussed in Section \ref{sec:conclusions}.

\section{Low-order models of shear flow}

We consider shear flow models in the spirit of the Galerkin models derived from the Navier-Stokes equations (see e.g. \cite{beneitez2020lcs} for a recent review). The generic structure of such models is a simple ordinary differential equation of the form
\begin{eqnarray}
\dot{{\bm x}}={\bm f}({\bm x})={\bm L}{\bm x}+{\bm N}({\bm x}),
\label{eq:modelsGeneral}
\end{eqnarray}
where ${\bm x} \in \mathbb{R}^n$, being $n$ a small integer, contains the relevant degrees of freedom of the system. ${\bm L}$ is the linear operator corresponding to the system linearized around the origin ${\bm x}={\bm 0}$, and $\bm{N}(\bm{x})$ represents the nonlinear terms. The model is such that the origin ${\bm x}={\bm 0}$ corresponds to the laminar fixed point $L$, while the turbulent state is simply represented by a different (attracting) fixed point $T$.
As pointed out in Ref.\ \cite{waleffe1997self}, for a model of subcritical transition to be consistent with the original PDEs it needs to be subjected to two constraints: (i) ${\bm L}$ is a linearly stable operator (i.e. all its eigenvalues have real negative parts) which allows for non-normal growth (ii) ${\bm{N}(\bm x)}$ conserves the total energy of the system, i.e. $\langle {\bm N}({\bm x}), \bm{x} \rangle = 0$ for all ${\bm x}$, where $\langle \cdot,\cdot\rangle$ is the scalar product.

The numerical solutions of the ODEs presented in this work have been obtained using Julia \cite{Julia-2017} with the package DifferentialEquations.jl \cite{rackauckas2017differentialequations}. The solver used is $\mathtt{Tsit5()}$ with relative and absolute tolerances of $10^{-14}$.

\subsection{Dauchot-Manneville model}
\label{sec:DM2D}

\begin{figure}
    \centering
    \includegraphics[width=1\columnwidth]{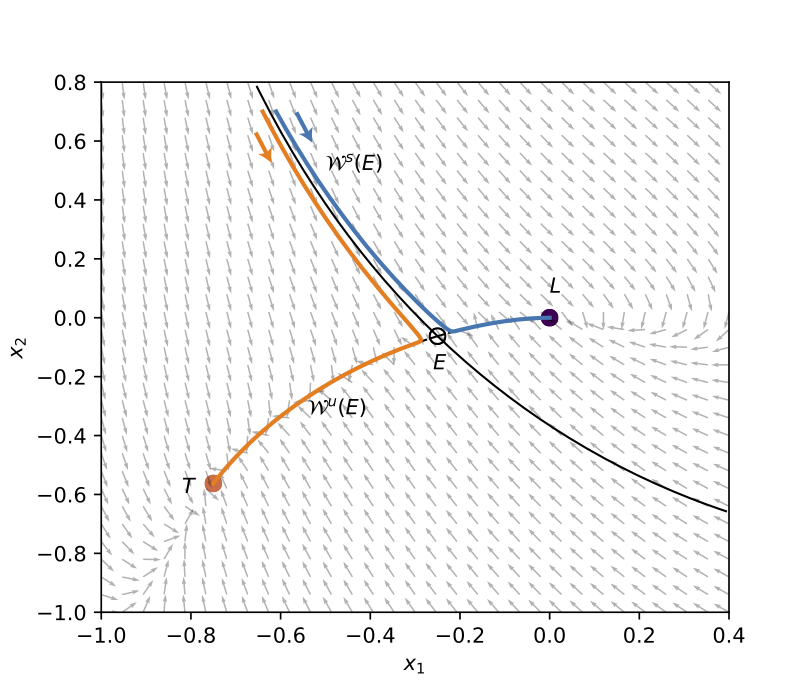}
    \caption{Phase portrait the 2D Dauchot and Manneville model. It is also the state portrait of the 3D model (independently of the value of $s_3$) restricted to the invariant plane $\mathcal{P}$ defined by $x_3=0$. Two trajectories approaching T and L are shown in orange resp. blue.}
    \label{fig:DM3D_2D_stateSpace}
\end{figure}

\begin{figure*}
    \centering
    \includegraphics[width=0.82\textwidth]{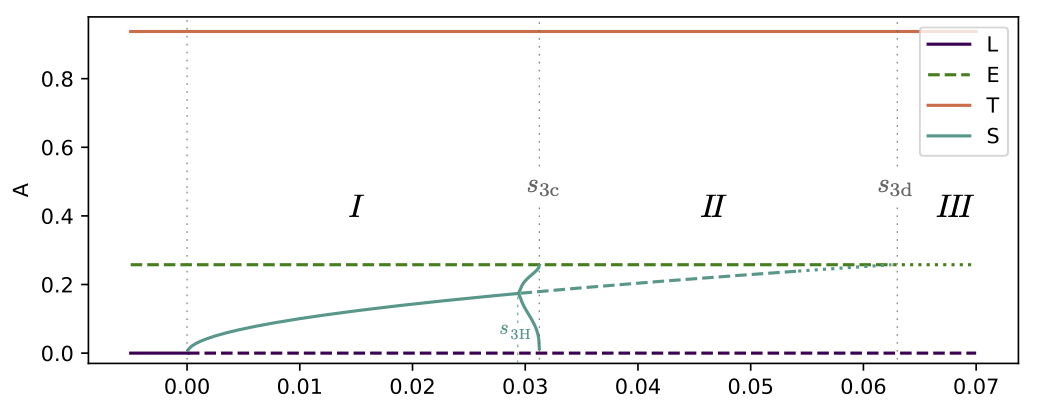} \put(-408,162){$(a)$}\\
    \includegraphics[width=0.82\textwidth]{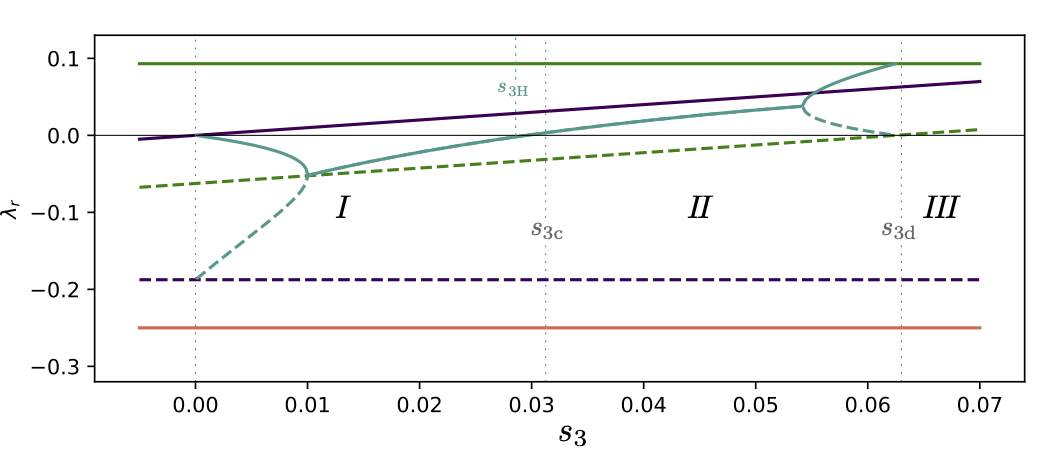} \put(-410,175){$(b)$}
    \caption{(a) Bifurcation diagram for the autonomous DM3D model. Amplitude $A$ of the steady and periodic solutions (minimum and maximum value only) vs. $s_3$. Stable solutions (solid lines), solutions with one (dashed lines) or two unstable eigenvalues (dotted lines). The limit cycle $\mathcal{C}$ appears at $s_3=s_{3H}\approx 0.03$ and collapses at $s_3=s_{3c}\approx 0.032$ (b) Real part of the eigenvalues of the linearized operator around each fixed point of the system versus $s_3$. The eigenvalues corresponding to perturbations of L,E and T within the plane $\mathcal{P}$ do not evolve with $s_3$. The fixed point $S$ arises as $L$, loses its linear stability at $s_3=0$, and disappears by merging with $E$ at $s_3=s_{3d}\approx 0.063$. Largest (solid) and second largest eigenvalue (dashed). Eigenvalues with real part $\leq -0.35$ are not shown. Vertical grey lines delimit the regions $I$, $II$ and $III$. The regions and parameters $s_{3H}$, $s_{3c}$ and $s_{3d}$  are described in the text}
    \label{fig:DM3D_bifurcationDiagram}
\end{figure*}
One of the simplest Galerkin models of shear flows was introduced by Dauchot and Manneville in Ref.\ \cite{Dauchot_1997}. The model (hereafter referred to as DM2D) is autonomous and assumes a two-dimensional state space. The equations for the DM2D model read
\begin{eqnarray}
\frac{dx_1}{dt} &=& s_1x_1+x_2 + x_1x_2\\
\frac{dx_2}{dt} &=& s_2x_2 - x_1^2,
\label{eq:DM2D}
\end{eqnarray}
where $s_1<0$ and $s_2<0$ are two real parameters interpretable as decay rates. The corresponding operator ${\bm L}$ is non-normal and definite negative. The effective control parameter in this two-dimensional model is the discriminant $\Delta=1-4s_1s_2$. We restrain the analysis to the range $\Delta <1$. For $\Delta < 0$ there is only one fixed point $L=(0,0)$. Two additional fixed points called $E$ and $T$ appear in a saddle-node bifurcation at $\Delta=0$. They are respectively unstable and stable and their expressions are
\begin{equation}
    E =\left( \frac{1}{2}(-1 + \sqrt{\Delta}),\frac{1}{4s_2}(-1 + \sqrt{\Delta})^2\right),
\end{equation}
and
\begin{equation}
    T =\left( \frac{1}{2}(-1 - \sqrt{\Delta}),\frac{1}{4s_2}(-1 - \sqrt{\Delta})^2 \right).
\end{equation}

The system DM2D is bistable with two well-defined basins of attraction in the interval $0\leq \Delta < 1$. The saddle point $E$ is the edge state of the system, whereas $L$ and $T$ are attractors. The basins of attraction of $L$ and $T$ are separated by a smooth edge manifold $\mathcal{W}^s(E)$ of the strong type according to the typology in Ref.\ \cite{lebovitz2012boundary}. A phase portrait for $s_1=-0.1875$ and $s_2=-1$ is shown in Fig.\ \ref{fig:DM3D_2D_stateSpace}.  The parameters $s_i,\ i=1,2$ can be re-interpreted as $s_i=-k_i^2/R$, where $-k_i^2,\ i=1,2$ arise from a Laplacian in Fourier space, and $R$ is a parameter akin to the Reynolds number in hydrodynamics. This phase portrait is ubiquitous in many fields of physics and results particularly convenient since only fixed points are present, while no chaotic dynamics take place.

\subsection{The autonomous 3D model}
\label{sec:DM3D}
We extend the DM2D by introducing a third variable $x_3$ in the model, orthogonal to $x_1$ and $x_2$. The extension is such that: (i) the laminar state $L$ will be unstable along this new direction (ii) the model still obeys the constraints on ${\bm L}$ and ${\bm N}(\bm x)$ and (iii) the dynamics within the 2D plane $\mathcal{P}$ defined by $x_3$=0 stays unchanged, i.e. $\mathcal{P}$ is an invariant plane for the new dynamics. The extended model reads
\begin{eqnarray}
\frac{dx_1}{dt} &=& s_1x_1+x_2 + x_1x_2\\
\frac{dx_2}{dt} &=& s_2x_2 - x_1^2 + \sigma x_3^2\\
\frac{dx_3}{dt} &=& s_3x_3  - \sigma x_2x_3,
\label{eq:DM3D}
\end{eqnarray}
where $s_3$ and $\sigma$ are two additional parameters of the model. The three-dimensional system is symmetric with respect to $\mathcal{P}$, which is an invariant plane and cannot be crossed by trajectories, so that we need only focus on the dynamics in the half-space $x_3\ge 0$. The case for which the model is investigated is $s_1=-0.1875$, $s_2=-1$ and $\sigma=-1$. The control parameter for this study is $s_3$. By construction, the dynamics restrained to $\mathcal{P}$ is unaltered compared to the 2D model even as $s_3$ varies, and is completely determined by the state portrait of Fig.\ \ref{fig:DM3D_2D_stateSpace}. However the stability of $L$, $E$ and $T$ in the direction $x_3$ transverse to $\mathcal{P}$ is now dependent on the value of $s_3$.

\subsection{Parametric study}
\label{sec:parStudy}
We explore the dynamics of the DM3D model by performing a parametric study in $s_3$ with a focus on the topological changes of the state space and the properties of the edge manifold for the half-space $x_3\ge 0$. The amplitude $A=\norm{\bm x}$ (with $\norm{\cdot}$ the Euclidian norm) of all fixed points and limit cycles is shown in a bifurcation diagram with respect to $s_3$ in Fig.\ \ref{fig:DM3D_bifurcationDiagram}(a). The real parts of the eigenvalues $\lambda_r$ associated with each fixed point of the system are shown in Fig.\ \ref{fig:DM3D_bifurcationDiagram}(b) with the convention that a positive real part indicates instability. They are used to assess the type of bifurcation undergone by the steady states of Fig.\ \ref{fig:DM3D_bifurcationDiagram}(a).
\begin{figure}
    \centering
    \includegraphics[width=0.85\columnwidth]{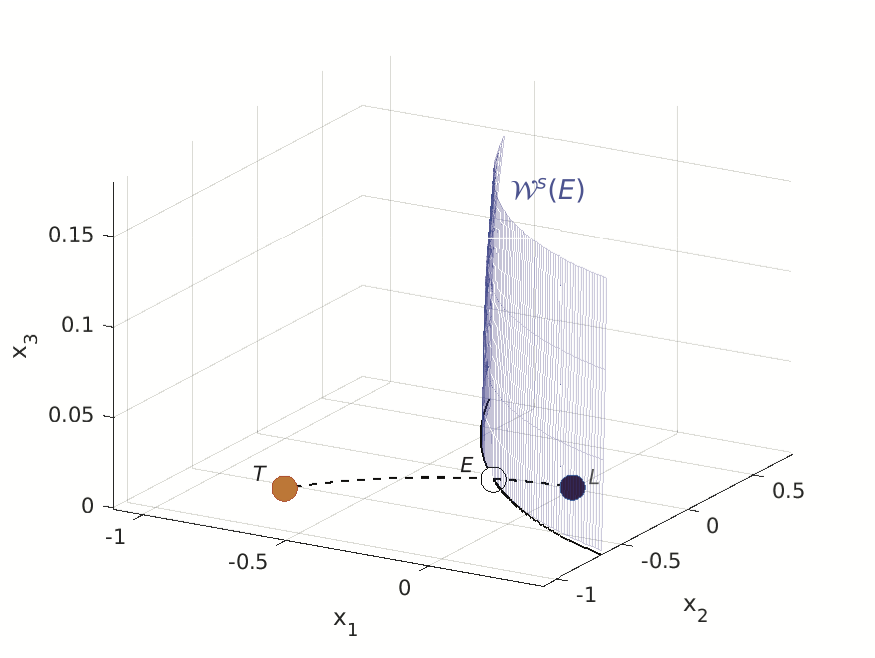} \put(-210,140){$(a)$}\\
    \includegraphics[width=0.85\columnwidth]{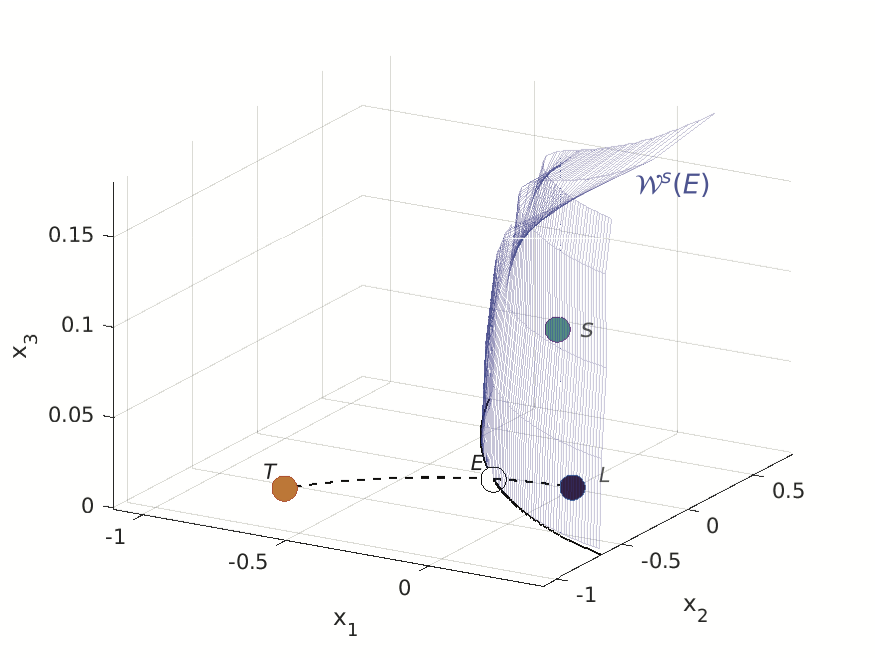} \put(-210,140){$(b)$}
    \caption{Autonomous DM3D. Comparison between region $s_3<0$ and $I$. Phase portraits for different values of $s_3$: (a) $s_3=-0.1$, (b) $s_3=0.01$. Surfaces (blue) correspond to the edge manifold $\mathcal{W}^s(E)$ identified using the method described in Ref.\ \cite{beneitez2020lcs}, Sec. II B.}
    \label{fig:3DstateSpace}
\end{figure}

The bifurcation diagram starts at $s_3<0$, for which the model has the same three fixed points, $L, T$ and $E$ as the original DM2D model. The only fixed point with an eigenvalue with a positive real part is the saddle point $E$. The associated three-dimensional state space is shown in Fig.\ \ref{fig:3DstateSpace}(a) for $s_3=-0.1$. Here, the two stable fixed  points $L$ and $T$ possess each their basin of attraction, respectively $\mathcal{B}(L)$ and $\mathcal{B}(T)$. The intersection of their closures, the edge manifold, coincides exactly with $\mathcal{W}^s(E)$, thus
\begin{equation}
\mathcal{W}^s(E)=\overline{\mathcal{B}(L)} \cap \overline{\mathcal{B}(T)}.
\end{equation}
$\mathcal{W}^s(E)$ is for $s_3<0$ is a two-dimensional surface invariant along $x_3$. It separates the basins of attraction of $L$ and $T$ and is then a codimension one manifold of the strong type.

For $s_3>0$ the laminar fixed point $L$ becomes linearly unstable, as the real part of its largest eigenvalue crosses $0$. For small enough $s_3>0$ the trajectories starting within neighbourhood of $L$ are attracted towards a new fixed point $S$. This new fixed point emerges by construction in a supercritical pitchfork bifurcation at $s_3=0$ (we focus only on the branch with $x_3$>0). The stability of $E$ and $T$ in the new state space remains unchanged, as shown in Fig.\ \ref{fig:DM3D_bifurcationDiagram}(b). $\mathcal{W}^s(E)$ is no longer invariant along $x_3$ and starts to curl around $S$, illustrated in Fig.\ \ref{fig:3DstateSpace}(b). However the two  basins of attraction of $T$ and $S$ (no longer $L$) exist and are separated by the edge manifold, which still coincides with $\mathcal{W}^s(E)$.

For $s_3 \gtrsim 0.01$, S is still stable but its leading eigenvalues are oscillatory: trajectories leaving $L$ now spiral in towards $S$. For $s_3=s_{3H} \approx 0.0296$, $S$ becomes unstable in a supercritical Hopf bifurcation, through which a stable limit cycle $\mathcal{C}$ is created. This limit cycle grows in amplitude until $s_3=s_{3c} \approx 0.03125$, where a global bifurcation takes place. The state space just before and after $s_{3c}$ is displayed in Fig.\ \ref{fig:3Dglobal_bif} and the global bifurcation is analysed in more detail in Section \ref{sec:globalbif}.
The state space continues to evolve beyond this global bifurcation. For $s_3=0.055$ the couple of complex conjugate eigenvalues of $S$ become two distinct real positive eigenvalues. For $s_3=s_{3d}=0.063$ the additional fixed point $S$ which was created at $s_3=0$ merges with the saddle point $E$ in a saddle-node bifurcation. Beyond $s_3=0.063$ the fixed points $L$ and $T$ are again the only attractors of the system. The saddle point $E$ still exists but has now two unstable eigenvalues as revealed by Fig.\ \ref{fig:DM3D_bifurcationDiagram}(b). All trajectories starting outside $\mathcal{P}$ converge towards $T$. The investigation has been pushed to $s_3=0.07$ without obvious change of state space topology.

\subsection{Global bifurcation of the state space}
\label{sec:globalbif}

\begin{figure}
    \centering
    \includegraphics[width=0.85\columnwidth]{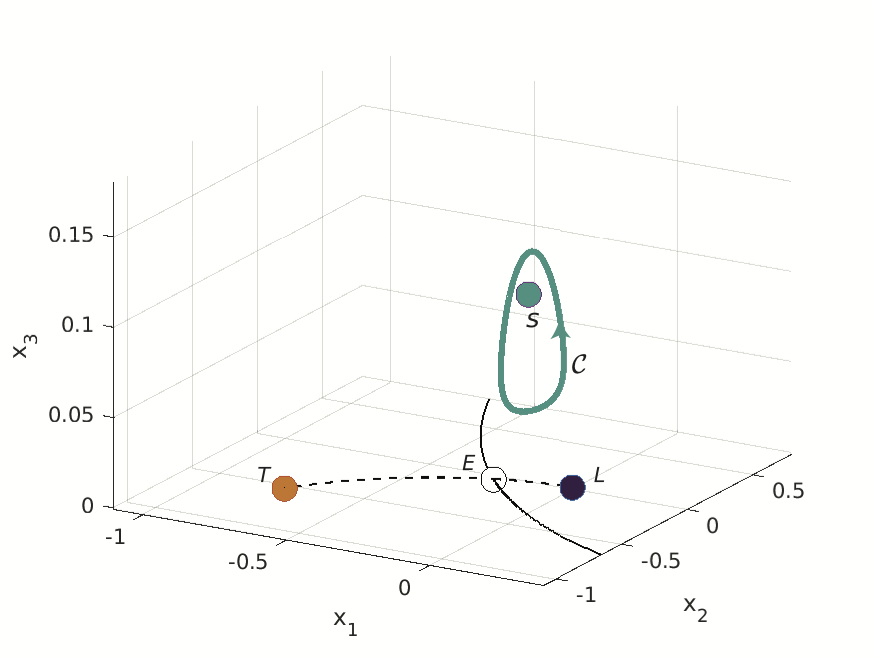} \put(-210,135){$(a)$}\\
    \includegraphics[width=0.85\columnwidth]{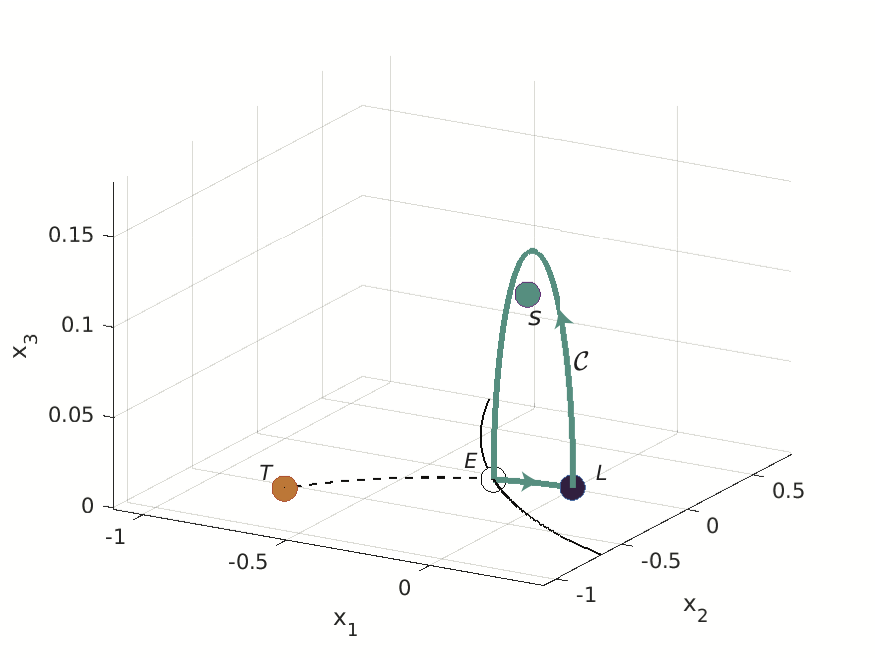} \put(-210,135){$(b)$}\\
    \includegraphics[width=0.85\columnwidth]{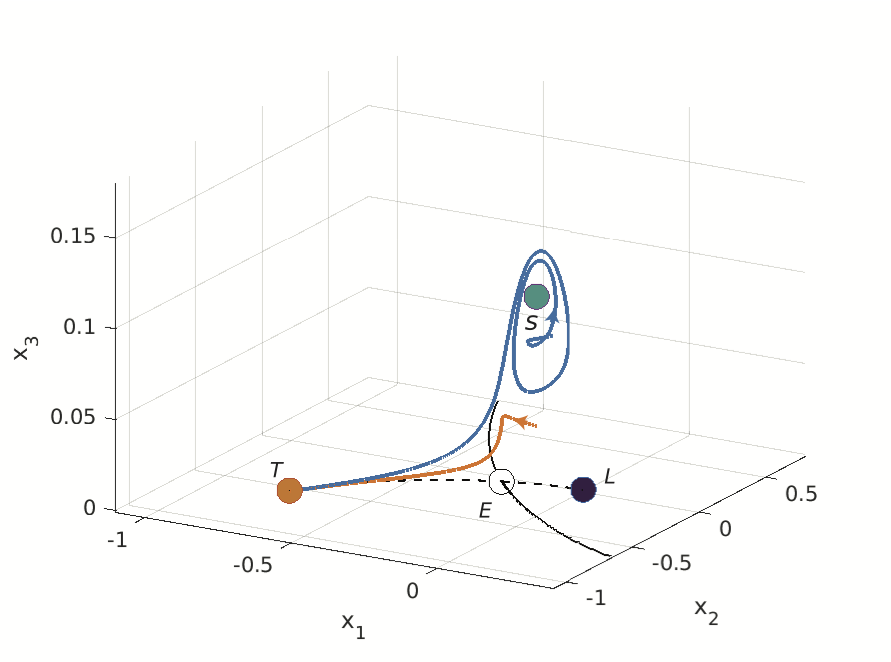} \put(-210,135){$(c)$}
    \caption{Autonomous DM3D. Comparison between region $I$ and $II$. Phase portraits for different values of $s_3$ close to the global bifurcation: (a) $s_3=0.031$. The system presents a hard edge between $\mathcal{C}$ and $T$, region $I$ (b) $s_3=0.03125$. Heteroclinic cycle close to $s_{3c}$ (c) $s_3=0.032$, beyond $s_{3c}$ the heteroclinic connection collapses and the edge $\mathcal{W}^s(E)$ becomes of the weak kind, region $II$. The blue and orange trajectories illustrate the two newly born different routes to transition.}
    \label{fig:3Dglobal_bif}
\end{figure}

\begin{figure}
    \centering
    \includegraphics[width=0.92\columnwidth]{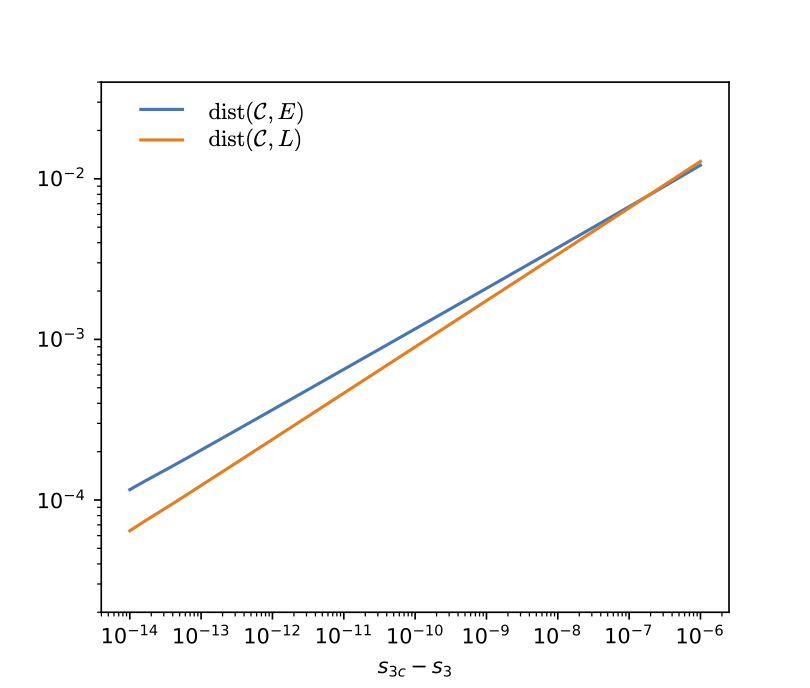}\put(-220,175){$(a)$}\\
    \includegraphics[width=0.92\columnwidth]{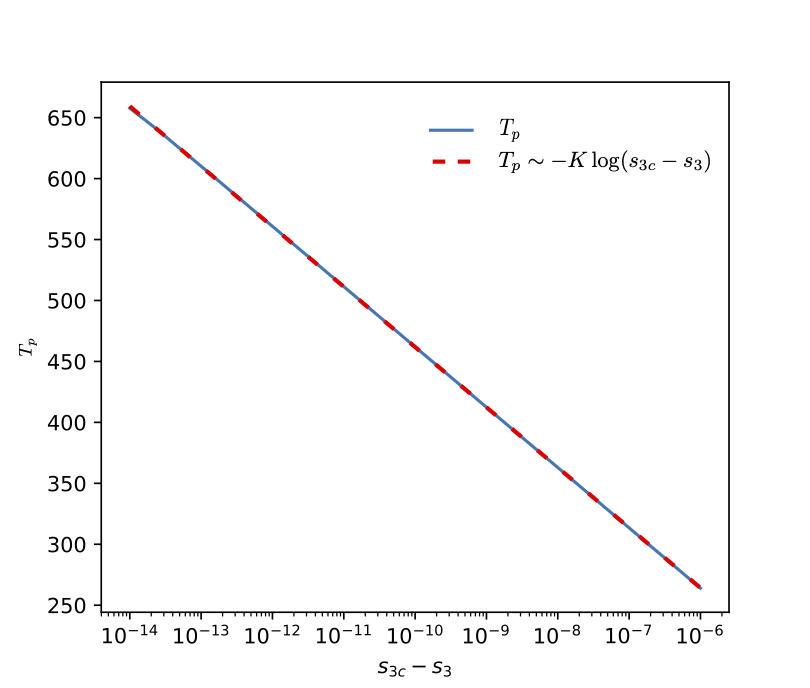}\put(-220,175){$(b)$}
    \caption{(a) Minimum distance from $E$/$L$-state to $\mathcal{C}$ as a function of the distance to criticality (b) Period $T_p$ of the limit cycle $\mathcal{C}$ versus the distance to criticality (solid), logarithmic fit with slope $K=49.4$ (dashed). The critical $s_{3c}\approx 0.03125$.}
    \label{fig:tau_vs_dist}
\end{figure}

This subsection is devoted to a more detailed analysis of the global bifurcation taking place at $s_{3c}\approx 0.03125$, because of its importance for the state space topology. Before the bifurcation, the limit cycle $\mathcal{C}$ grows in amplitude, as shown in Fig.\ \ref{fig:tau_vs_dist}(a), the minimum distances between $\mathcal{C}$ and, both $E$ and $L$, decreases as $s_3$ approaches $s_{3c}$. The two distances both scale algebraically with $\norm{s_3-s_{3c}}$ and, importantly,
these two distances reach zero at the same value of $s_3=s_{3c}$ (determined to numerical accuracy using double precision arithmetic). The period of the limit cycle $\mathcal{C}$ is also monitored as a function of $\norm{s_3-s_{3c}}$ in Fig.\ \ref{fig:tau_vs_dist}(b): a logarithmic fit of the form $T_p \sim \log(|s_3-s_{3c}|)$ is valid over as much as eight decades. This is an unambiguous signature for a saddle-loop collision \cite{gaspard1990measurement, strogatz2001nonlinear}. Note that compared to \cite{gaspard1990measurement} where the collision takes place between an attracting limit cycle and a saddle point, in the present case $\mathcal{C}$ collides simultaneously with two saddle points (however the scaling for the diverging period $T_p$ does not change).

At the bifurcation point, $E$ and $L$ form a heteroclinic network: the heteroclinic connection from $E$ to $L$ lies in the invariant plane $\mathcal{P}$ whereas the connection from $L$ to $E$ lies outside $\mathcal{P}$. The bifurcation occurs as $\mathcal{C}$ collides with the heteroclinic cycle connecting $E$ to $L$. Close to criticality $\mathcal{C}$ deforms close to the fixed points, as seen in Fig.\
\ref{fig:3Dglobal_bif}(b).

Just before the global bifurcation takes place, there are still two basins of attraction, respectively $\mathcal{B}(T)$ and $\mathcal{B}(\mathcal{C})$. The intersection of their closures, the edge manifold, coincides with $\mathcal{W}^s(E)$. However, once the heteroclinic connection is formed, the manifold separating the two different basins of attraction becomes of the weak type, as illustrated by its state space in Fig.\ \ref{fig:3Dglobal_bif}(c), where trajectories following the two newly born routes to transition are shown. The global bifurcation changes the whole structure of the state space in the following way: for $s_3>s_{3c}$ the two disjoint different basins of attraction no longer exist as open sets. All trajectories starting outside $\mathcal{P}$ asymptotically approach $T$ in forward time.
Note that trajectories starting strictly in $\mathcal{P}$ can still reach $L$ provided they belong to the relevant part (which coincides with the embedding in $\mathbb{R}^3$ of the basin $\mathcal{B}(L)$ from DM2D), however $\mathcal{P}$ is no longer an open set in $\mathbb{R}^3$ and has measure zero.

 A useful measure to understand the state space is the transition time $\tau_{tr}$, introduced in \cite{zammert2019transition} as the time it takes for a trajectory to approach the turbulent attractor within a given pre-determined (small) distance $\epsilon_T$, as a function of the starting point ${\bm x}$. Mathematically it is defined as $\argmin_{t>0} \norm{ \phi^t({\bm x})-T }<\epsilon_T$, where $\phi^t$ refers to the flow map. Iso-$x_2=-0.2$ cross-sections of the transition time for $s_3 \lesssim s_{3c}$ and $s_3 \gtrsim s_{3c}$ shown in Fig.\ \ref{fig:LTmaps}(a)-(b) respectively confirm that only the basin of attraction $\mathcal{B}(T)$ is left. The stable manifold of E, $\mathcal{W}^s(E)$, still exists as a codimension one manifold except that it appears now winded around $S$. It is now possible to distinguish a faster and a slower route to turbulence, depending on which side of $\mathcal{W}^s(E)$ the initial condition is located. As a consequence the edge manifold is no longer interpretable as a basin boundary any more, but rather as a soft boundary in state space between two routes to transition.

\subsection{Accessibility of the edge state}
\label{sec:accessibility}
The interpretation of the edge manifold, beyond criticality, as a manifold separating two different routes, one fast and one slow, to the same attractor holds for the parametric region of interest after the global bifurcation. There exists however a fundamental difference between two behaviours, for $s_3\in(0.03125, 0.063)$ and $s_3>s_{3d}=0.063$, linked to the notion of \emph{accessibility}  similar to that used in Ref.\ \cite{grebogi1987basin}.

\begin{figure}
    \centering
    \includegraphics[width=0.92\columnwidth]{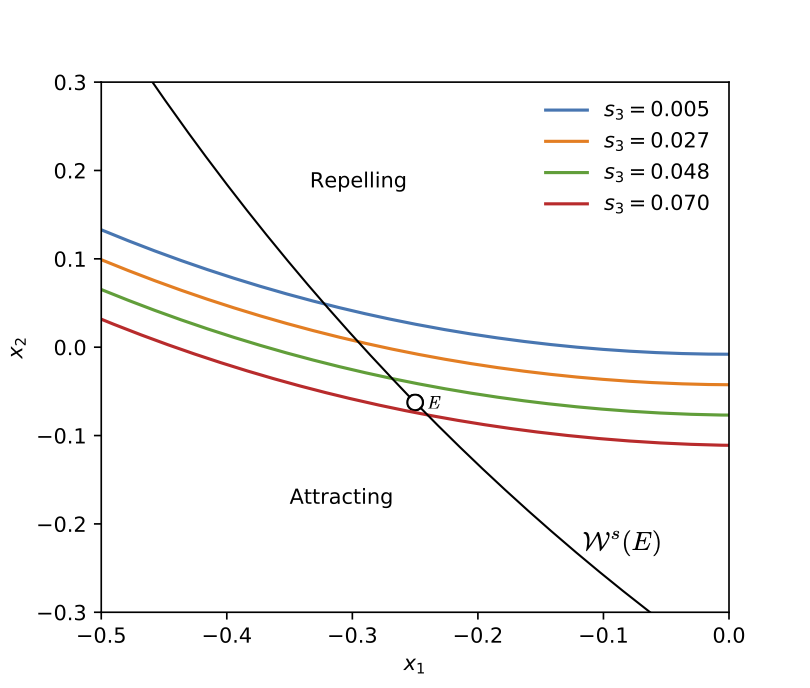}
    \caption{Transverse FTLE at $x_3$=0 for a  time horizon $\tau=1$. Solid black line indicates $\mathcal{W}^s(E)$. Color lines indicate the $0$-isovalue of the transverse FTLE for different $s_3$.}
    \label{FTLEs}
\end{figure}

\begin{figure}
    \centering
    \includegraphics[width=0.9\columnwidth]{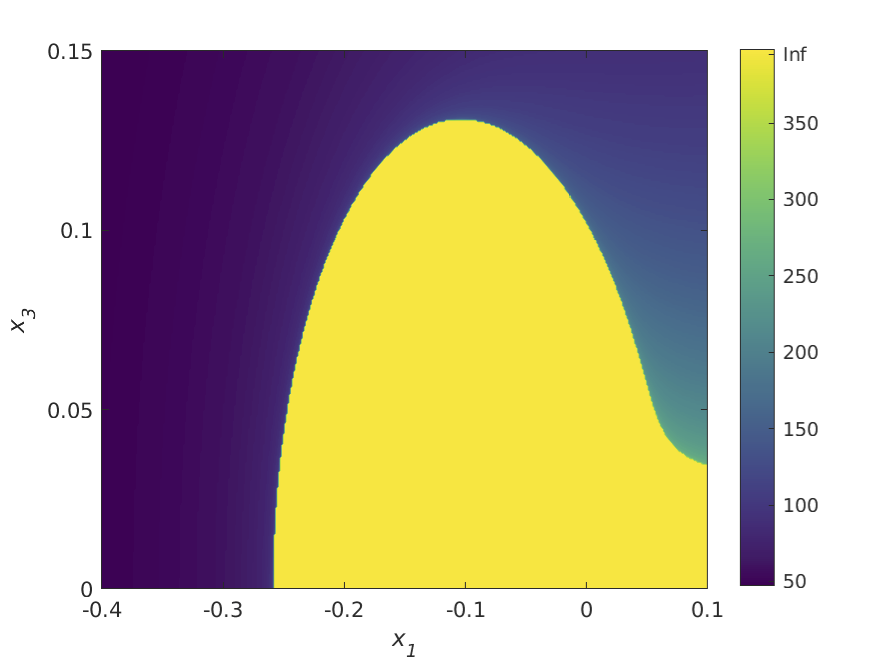}\put(-222,157){$(a)$}\\
    \includegraphics[width=0.9\columnwidth]{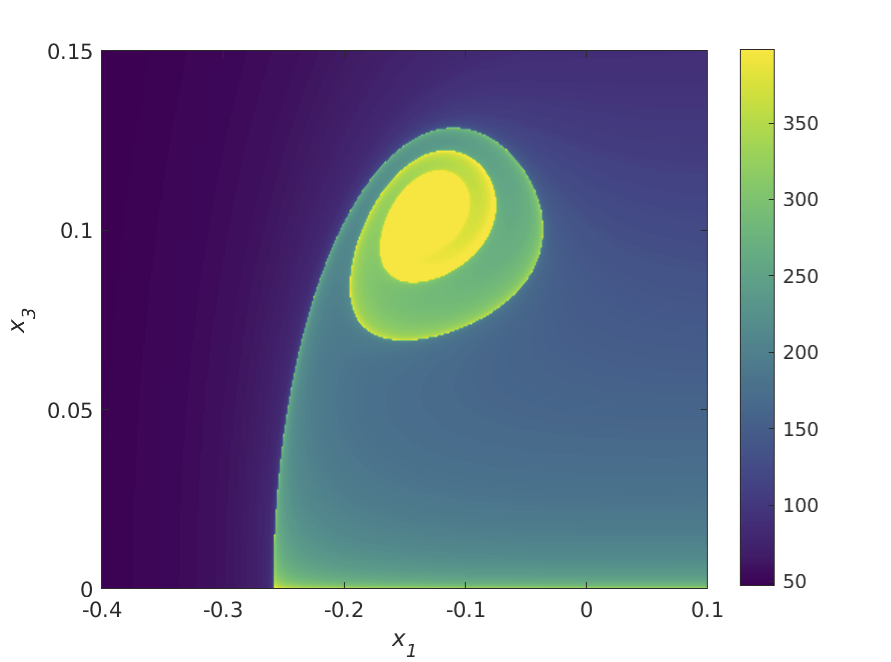}\put(-222,157){$(b)$}\\
     \includegraphics[width=0.9\columnwidth]{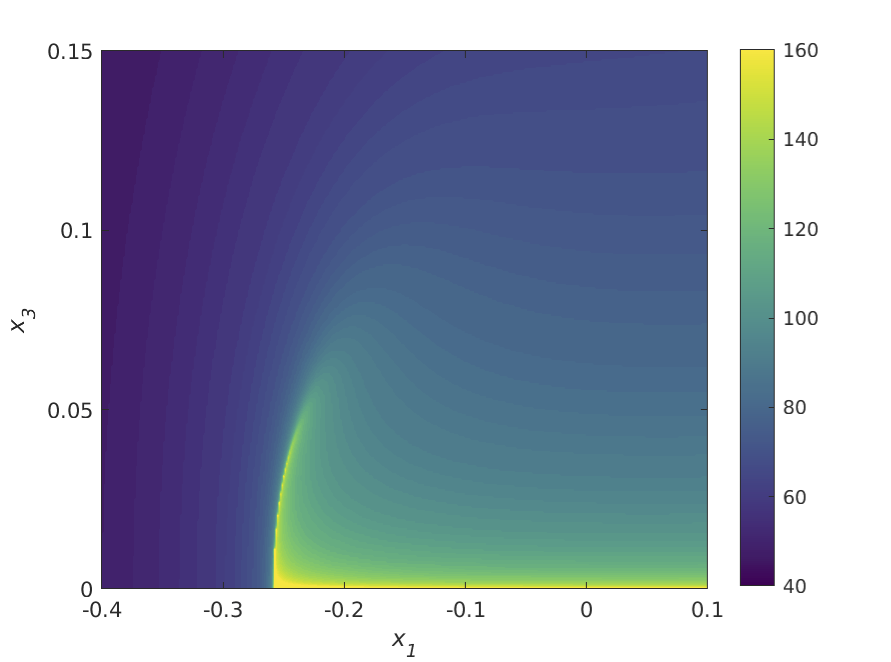}\put(-222,157){$(c)$}
    \caption{Transition times $\tau_{tr}$ for $x_2x_3$-plane for $x_1=-0.2$ (a) $s_3=0.031249$, within region $I$ (b) $s_3=0.032$, within region $II$ and (c) $s_3=0.065$, within region $III$.}
    \label{fig:LTmaps}
\end{figure}

\begin{figure*}
    \centering
    \includegraphics[width=0.45\textwidth]{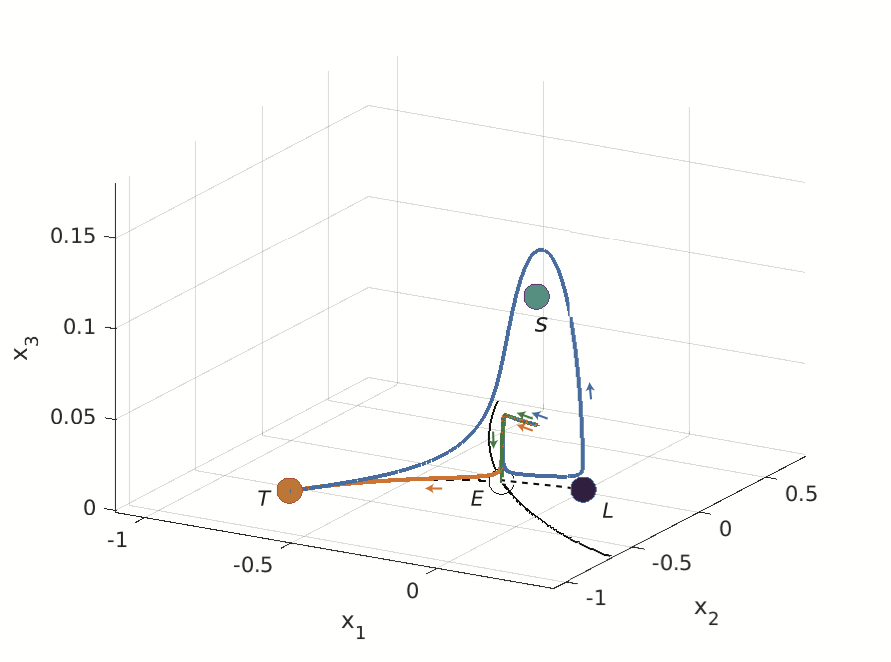} \put(-220,145){$(a)$}
    \includegraphics[width=0.45\textwidth]{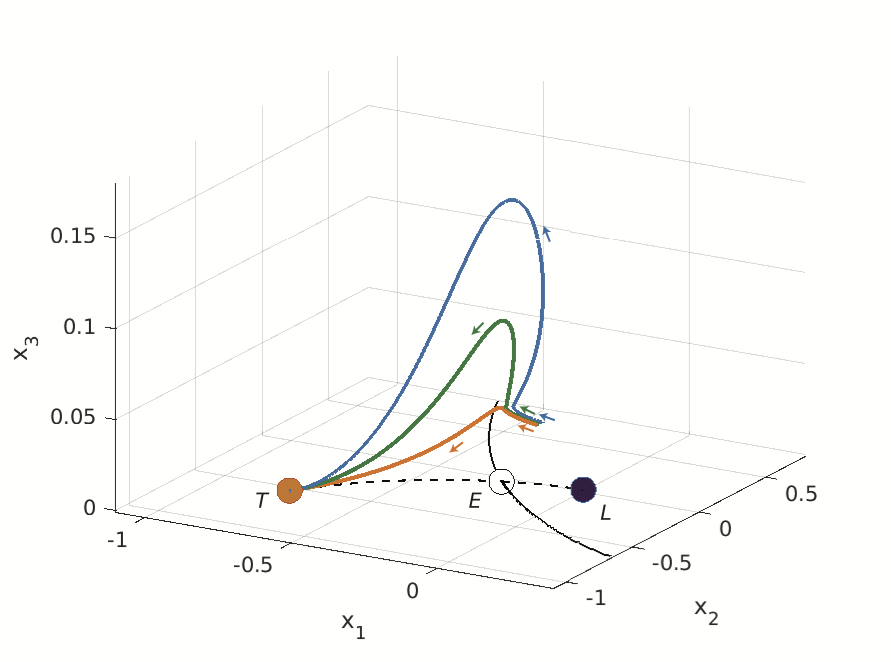} \put(-220,145){$(b)$} \\
    \includegraphics[width=0.45\textwidth]{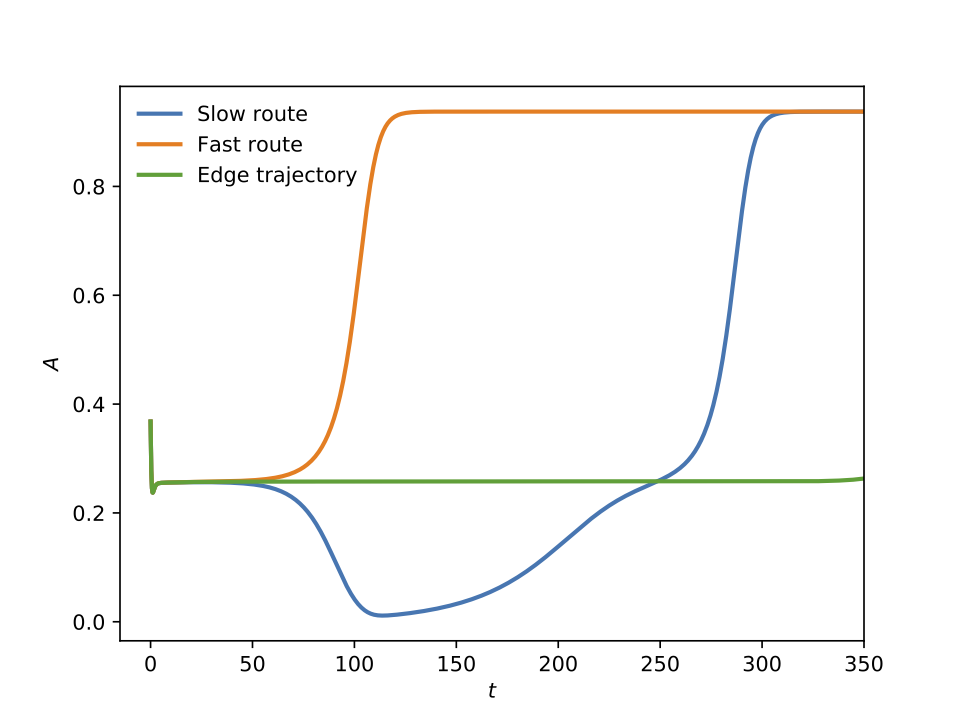} \put(-220,155){$(c)$}
    \includegraphics[width=0.45\textwidth]{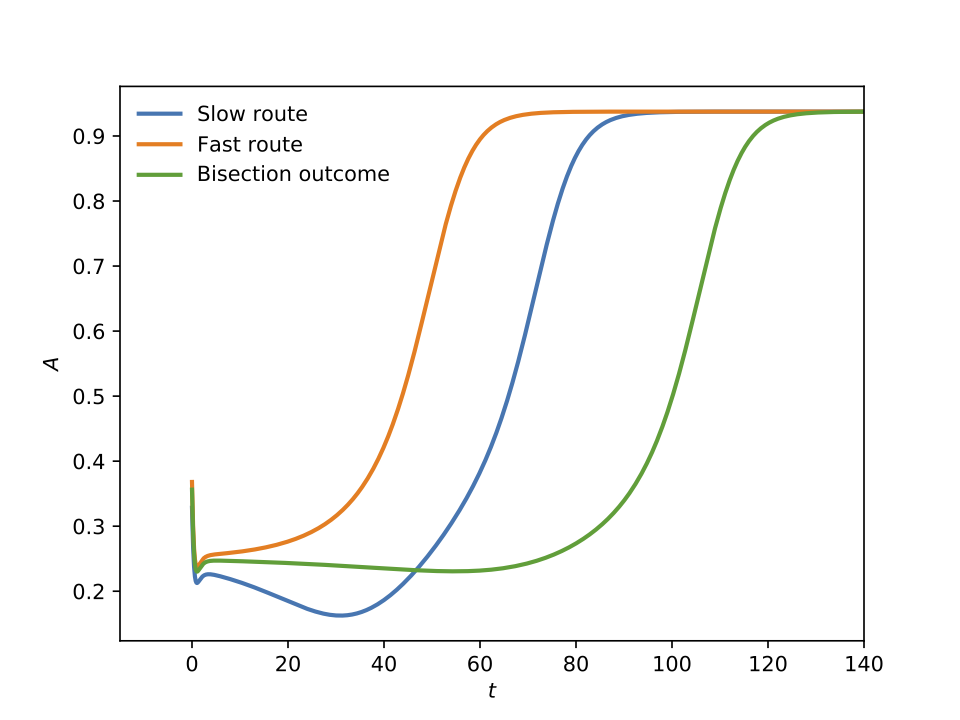}\put(-220,155){$(d)$}
    \caption{Comparison between region $II$ and $III$. Trajectories on different sides of the edge manifold starting from initial conditions with $x_3=0.05$. (a) $s_3=0.032$, the edge state is accessible by bisection from outside $\mathcal{P}$, region $II$. (b) $s_3=0.065$, the edge state is no longer accessible by bisection  from outside $\mathcal{P}$, region $III$. (c) time series for $A(t)$ for the trajectories in (a). (d) idem for the trajectories in (b). The blue and orange trajectories illustrate the two newly born different routes to transition, to be compared with the same colours in Fig.\ \ref{fig:blasius}(b).}
    \label{fig:DM3D_exampleAmps}
\end{figure*}

For low $s_3<s_{3d}$, $E$ has only one unstable eigenvalue, and its unstable manifold is fully contained in $\mathcal{P}$. This behaviour is illustrated in the state portrait in Fig \ref{fig:DM3D_exampleAmps}(a). It is hence possible to approach $E$ using a trajectory starting outside $\mathcal{P}$ and $E$ is labelled as \emph{accessible}. This is shown in the time series of three trajectories with closeby initial conditions in Fig.\ \ref{fig:DM3D_exampleAmps}(c), where the edge trajectory approaches $E$ starting from an initial condition with $x_3(0)\neq 0$ and remains in its neighbourhood (at least transiently). As a consequence the trajectories following the slow route to turbulence close to the edge manifold approach $L$ more closely than those away from $E$'s neighbourhood.

However, for $s_3>0.063$ $E$ has two distinct unstable eigenvalues, and is now also repelling in the direction transverse to $\mathcal{P}$. Finding a trajectory converging towards $E$ from an initial condition outside $\mathcal{P}$ is now impossible. $E$ is said to be \emph{non-accessible}. This behaviour is illustrated using a phase portrait in Fig.\ \ref{fig:DM3D_exampleAmps}(b) and using the time series corresponding to three trajectories starting from nearby initial conditions in Fig.\ \ref{fig:DM3D_exampleAmps}(d): unlike in Fig.\ \ref{fig:DM3D_exampleAmps}(a) we observe now that the trajectory sandwiched between the fast and slow routes in state space does not approach $E$. Consequently, the trajectories close to the edge manifold which follow the slow route do not visit the neighbourhood of $E$.

The notion of accessibility of a given fixed point is here directly related to the codimension of its stable manifold. Here the saddle point $E$ is accessible when codim $(\mathcal{W}^s(E))$=1, in which case $E$ is an edge state, and non accessible when codim $(\mathcal{W}^s(E))$>1, in which case the system does not have any edge state. Note how this is dependent on dimensions: for any $s_3$ the system DM2D restrained to $\mathcal{P}$, which is of dimension 2 only, possesses $E$ as an edge state. A simple local diagnostic for the instability of $E$ in a direction transverse to the invariant plane $\mathcal{P}$, consists in computing the transverse finite-time Lyapunov exponent (FTLE) $\lambda_T$. For any ${\bm x} \in \mathcal{P}$, we define $P_3$ as the projection on $\mathcal{P}^{\perp}$ such that $P_3(x_1,x_2,x_3)=(0,0,x_3)$, and
\begin{eqnarray}
\lambda_T({\bm x},\tau)=\frac{1}{\tau}\log\left(\frac{\norm{P_{3}(\phi^{\tau}({\bm  x}))}}{\norm{P_{3}({\bm x})}}\right).
\end{eqnarray}
A point ${\bm x} \in \mathcal{P}$ such that $\lambda_T({\bm x},\tau)>0$ for $\tau$ sufficiently small is non accessible from outside $\mathcal{P}$. Iso-levels of transverse FTLEs are shown in Fig.\ \ref{FTLEs} for the horizon time $\tau$=1. By construction $\lambda_T$ is always positive at ${\bm x}=L$ as soon as $s_3>0$, whereas at ${\bm x}=T$ it is negative since $T$ is always stable. The locus where $\lambda_T$ crosses zero hence marks the stability boundary in $\mathcal{P}$ between trajectories initially deviating from
$\mathcal{P}$ and those attracted by $\mathcal{P}$. The iso-line $\lambda_T$=0 (for a given value of $\tau$) moves with changing $s_3$. From Fig. \ref{FTLEs} it is clear that $E$ lies on the negative side of the stability boundary for $s_3<s_{3d}$ and on the positive side for $s_3>s_{3d}$. $s_3=s_{3d}$ hence marks the loss of accessibility of $E$. This has strong implications for the bisection process and the numerical determination of the edge state. When $E$ is accessible, a one-dimension search (i.e. a bisection) along almost any line in state space will generate one trajectory converging to $E$ (even if this trajectory is repelling). When $E$ is not accessible this is no longer the case and the bisection algorithm is not warranted to converge from any initial condition: there is no edge state although $E$ still exists as a saddle point. Concretely, if the aim is to identify $E$ as an edge state, the only possibility is to apply a control strategy which leaves $E$ unchanged but affects its effective transverse stability. Any control strategy that can make $\lambda_T(E,\tau)$ negative for some value of $\tau$ is likely to make bisection algorithms converge to $E$.

The loss of accessibility of the edge state is further illustrated  in Fig.\ \ref{fig:LTmaps}(c) using transition times computed in a section with $x_2=-0.05$. The Figure highlights larger values of $\tau_{tr}$ in some regions of the state space, however the highlighted structure outside $\mathcal{P}$ diffuses out and is not a sharp boundary for the trajectories on either sides, in contrast to Fig.\ \ref{fig:LTmaps}(a) and (b).

\section{The nonautonomous 3D model}
\label{sec:DM3Dnonauto}

The autonomous model DM3D has the discriminant $\Delta$, or indirectly the Reynolds number $R$, as a control parameter as is the case for e.g. plane Poiseuille flow. In an effort to mimic the evolution of a localised disturbance in a spatially developing flow, one needs to take into account the fact that the disturbance experiences changing values of $R$ as it is advected downstream at an approximately constant velocity. The simplest way to introduce a time dependence in the DM3D model is to make the growth rate $s_3$ time-dependent, so that it becomes positive only at a finite time and continues to increase subsequently. This ensures that $L$, $E$ and $T$ are still fixed points of the system for every time. A linear relation of the type $s_3(t)=k_1t+k_2$ fulfills these requirements with as few parameters as possible. The system becomes then nonautonomous, while retaining all the properties required for models of subcritical transition. It reads
\begin{eqnarray}
\frac{dx_1}{dt} &=& s_1x_1+x_2 + x_1x_2\\
\frac{dx_2}{dt} &=& s_2x_2 - x_1^2 + \sigma x_3^2\\
\frac{dx_3}{dt} &=& (k_1 t+k_2)x_3  - \sigma x_2x_3.
\label{eq:DM3Dna}
\end{eqnarray}
The time interval is restrained to $t\in [t_0,t_0+T_F]$, with $k_1=0.73/T_F$ and $k_2=-0.1$, in order to facilitate the comparison with its autonomous counterpart.

\begin{figure}
    \centering
    \includegraphics[width=0.9\columnwidth]{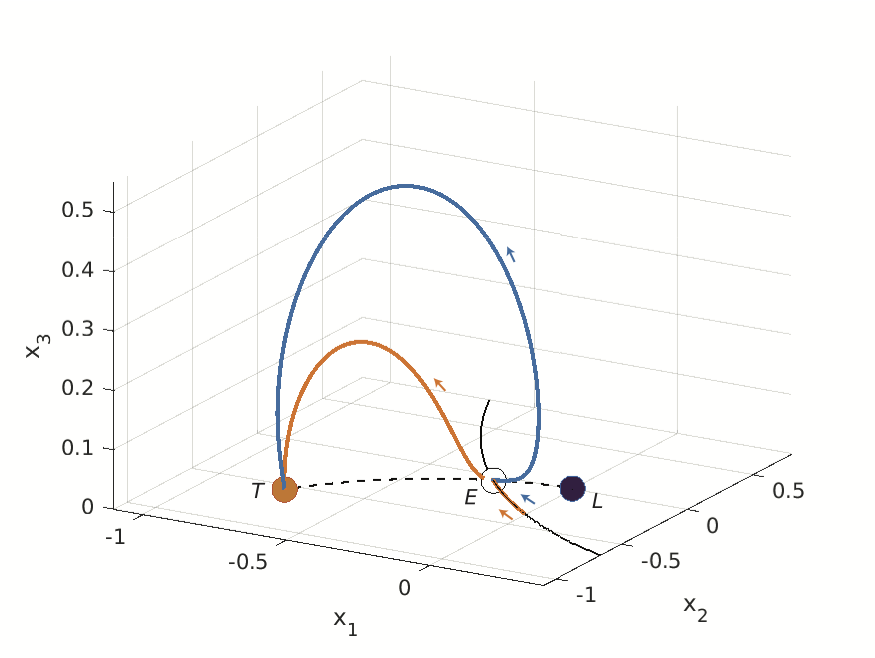} \put(-225,130){$(a)$}\\
    \includegraphics[width=0.9\columnwidth]{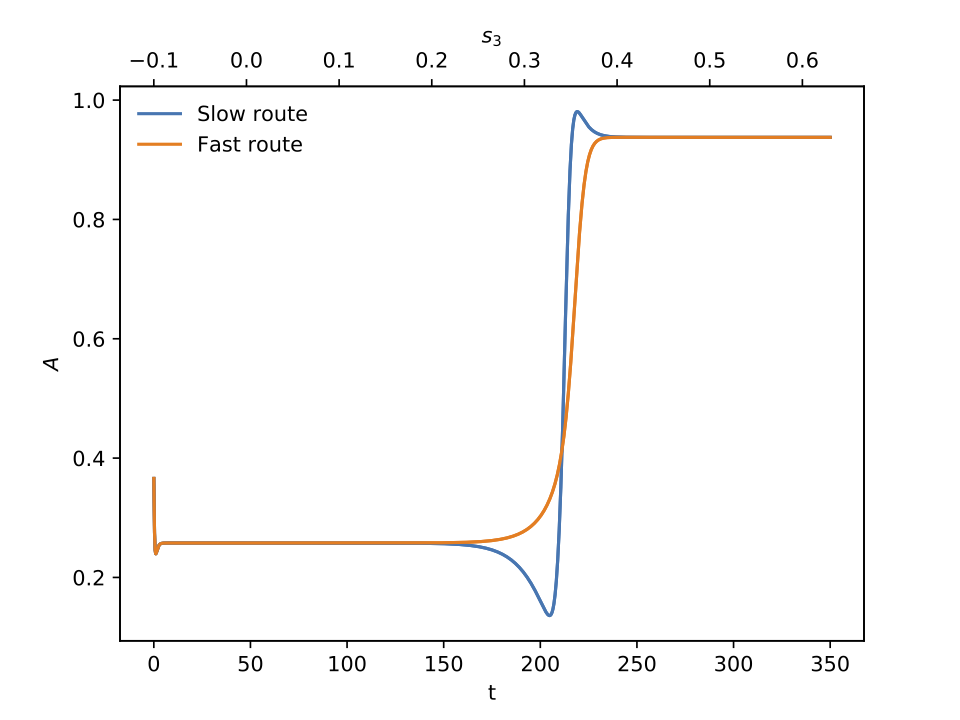} \put(-223,150){$(b)$}
    \caption{DM3D nonautonomous model. (a) state portrait for two trajectories starting close to the edge of DM2D and to the invariant plane  $\mathcal{P}$, (b) time series of the amplitudes $A(t)$ along the same trajectories trajectories}
    \label{fig:3D_nonauto}
\end{figure}

A state portrait of the nonautonomous system is shown for several trajectories with $t_0=0$ and $T_F=350$ in Fig.\ \ref{fig:3D_nonauto}(a). The chosen trajectories start very close to the edge manifold within $\mathcal{P}$ from an initial condition with $x_3(t_0)=10^{-6}$, with colour coding chosen to match Fig.\ \ref{fig:blasius} and Fig.\ \ref{fig:DM3D_exampleAmps}(a),(b). The time series for the amplitude along the trajectories are plotted in Fig.\ \ref{fig:3D_nonauto}(b).
The fixed point $T$ in DM3D can be reached in two different ways from outside the $x_1x_2$ plane, either approaching $L$ (slow route) or not (fast route), as shown in Fig.\ \ref{fig:3D_nonauto}(a). The model experiences thus a Blasius-like dynamics when considering trajectories starting at a weak non-zero distance from the invariant plane $\mathcal{P}$.

\section{Discussion}
\label{sec:conclusions}
In the present article we have devised and investigated two low-order models, one autonomous and the other one nonautonomous. These models aim at explaining non-standard behaviours observed during bisection in bistable systems once one of the attractors undergoes a bifurcation and becomes unstable \cite{beneitez2019edge,zammert2019transition,canton2020critical}. The present generalisation respects the constraints due to the non-normality of the linearised operator as well as the energy-conserving nonlinearity due to the Reynolds-Orr theorem \cite{schmid2001stability}. The procedure is not limited to the two-dimensional Dauchot-Manneville and can be carried out in higher dimension using reference models such as e.g. the four-dimensional Waleffe model \cite{waleffe1997self}, however the present three-dimensional is ideal for visualising state space boundaries.

The nonautonomous DM3D model has dynamics similar to DM2D as long as the new parameter $s_3<0$. As $s_3$ increases above zero, the laminar state $L$ becomes unstable in favour of a new fixed point $S$. After a series of local bifurcations, $S$ becomes a limit cycle $\mathcal{C}$. The first topological change occurs at $s_3=s_3c$ as $\mathcal{C}$ collides with the two saddle points $E$ and $L$: the stable manifold $W^s(E)$ still exists as a codimension one manifold but does not split the state space into two any longer, instead all trajectories outside the invariant plane $\mathcal{P}$ are attracted by the turbulent fixed point $T$. Past a second bifurcation at $s_3=s_{3d}$, $E$ gains one more unstable eigenvalue so that codim $(\mathcal{W}^s(E))$ goes from 1 to 2. $\mathcal{W}^s(E)$ is no longer the edge manifold and $E$ no longer the edge state because it has lost its accessibility property, and the edge has collapsed, i.e. there is no edge anymore. The only possibility to reach $E$ from outside $W^s(E)$ $\mathcal{P}$ is to use a control strategy in order to restabilise the laminar state $L$ locally.

The present extension of a low-order model for subcritical transition allows for a more generic investigation of bistable systems, when a linear instability disturbs the usual dynamical picture. It also paves the way for the study of the edge manifold beyond criticality in higher-dimensional systems, such as the full Navier-Stokes equations. In fluid systems there are several canonical examples where both transition scenarios are present, such as plane Poiseuille flow \cite{zammert2019transition}, the Blasius boundary layer \cite{beneitez2019edge} or bent pipe flow \cite{canton2020critical}. Although a detailed comparison of these flow cases with the present models would be naive, common features related to the global topology of thee state space are relevant. The bent pipe flow system investigated in Ref.\ \cite{canton2020critical} is the simplest case: it contains two attractors, like in any bistable system, except that a non-trivial (a travelling wave) has inherited the former stability of the laminar flow via a local supercritical bifurcation. Although the precise nature of the bifurcation differs, this situation is topologically similar to the autonomous model DM3D for $0<s_3<s_{3c}$: two basins of attraction separated by an strong edge manifold, including a limit cycle on one side. The plane Poiseuille flow configuration investigated in Ref.\ \cite{zammert2019transition} for $\textit{Re}=5855$ is more complex: the edge state solution still exists whereas there is only one attraction basin, namely the turbulent basin. This is analog to the state space picture of the autonomous model for $s_{3c}<s_3<s_{3d}$, where the edge manifold still exists as a codimension one hypersurface except that it does not split the state space into two disjoint regions any longer. Eventually, the case of the Blasius boundary layer, shown in Fig. \ref{fig:blasius}, is the most complex: although early time bisection seems to indicate that the edge state is accessible along a trajectory starting from a well-tuned initial condition, it is lost for larger times. The early and later times can be informally compared to the lower and higher $s_3>0$ regimes of the DM3D model, respectively below and above $s_{3d}$. The nonautonomous model, by construction, sweeps as time progresses through the same range of values of $s_3$ as its autonomous counterpart.
As shown in Fig.\ \ref{fig:3D_nonauto}, initial conditions close, yet outside the invariant plane $\mathcal{P}$, would first shadow the edge trajectory in $\mathcal{P}$. At a later time they would be ejected away from $\mathcal{P}$ and converge towards the turbulent set $T$ in two possible ways: either via a transient approach to $L$ (blue trajectories in Fig.\ \ref{fig:3D_nonauto}) or without it (orange trajectories in Fig.\ \ref{fig:3D_nonauto}). The edge manifold is only defined for short finite times but has no existence as an invariant set over unbounded times. The edge state is not accessible anymore except from initial conditions strictly inside $\mathcal{P}$, or at the cost of a control strategy. This confirms and explains the conclusions of Ref.\ \cite{beneitez2019edge} based on costly edge tracking of the Navier-Stokes equations in large computational domains (cf both their Fig. 10 and the present Fig. \ref{fig:blasius} compared to the current Fig.\ \ref{fig:3D_nonauto}).

The two models presented here are simple, non-chaotic and their three-dimensional state space has the important advantage of being easily visualisable. Yet they offer the possibility to understand the topology of seemingly hopelessly tangled high-dimensional state spaces associated with several fluid problems. We believe that the present strategy can be used in many diverse areas of physics where deterministic bistability is disturbed by an additional local bifurcation.

\begin{acknowledgments}
 Financial support by the Swedish Research Council (VR) grant no. 2016-03541 is gratefully acknowledged. The authors also thank Philipp Schlatter for discussions on the subject. The open source projects
\href{https://julialang.org/}{Julia},
\href{https://matplotlib.org/}{Matplotlib},
and
\href{https://www.paraview.org/}{ParaView}
have been used for this work.

\end{acknowledgments}

%%%%%%%%%% Insert bibliography here %%%%%%%%%%%%%%

\bibliography{main.bib}

\end{document}